\documentclass
[aps,prb,reprint,superscriptaddress,amssymb,amsmath,amsfonts,floatfix]{revtex4-1}
\usepackage{graphicx}
\usepackage{hyperref}
\usepackage{bm}

\AtBeginDocument{
 \newwrite\bibnotes
 \def\bibnotesext{Notes.bib}
 \immediate\openout\bibnotes=\jobname\bibnotesext
 \immediate\write\bibnotes{@CONTROL{REVTEX41Control}}
 \immediate\write\bibnotes{@CONTROL{
  apsrev41Control,author="08",editor="1",pages="1",title="0",year="0"}}
 \if@filesw
  \immediate\write\@auxout{\string\citation{apsrev41Control}}
 \fi
}

\newcommand{\SRO}{Sr\textsubscript{2}RuO\textsubscript{4}}

\begin{document}
\title{A key role of correlation effects in the Lifshitz transition in \texorpdfstring{\SRO{}}{Sr2RuO4}}
\author{Mark E. Barber}
\altaffiliation[Present address: ]{Department of Applied Physics and Geballe Laboratory for Advanced Materials, Stanford University, Stanford, CA 94305}
\email{mebarber@stanford.edu}
\affiliation{Max Planck Institute for Chemical Physics of Solids, N{\"o}thnitzer Stra{\ss}e 40, 01187 Dresden, Germany}
\author{Frank Lechermann}
\affiliation{I.\ Institut f{\"u}r Theoretische Physik, Universit{\"a}t Hamburg, Jungiusstra{\ss}e 9, 20355 Hamburg, Germany}
\author{Sergey V. Streltsov}
\author{Sergey L. Skornyakov}
\affiliation{Institute of Metal Physics, S.\ Kovalevskaya St.\ 18, 620990 Ekaterinburg, Russia}
\affiliation{Department of Theoretical Physics and Applied Mathematics, Ural Federal University, Mira St.\ 19, 620002 Ekaterinburg, Russia}
\author{Sayak Ghosh}
\author{B. J. Ramshaw}
\affiliation{Laboratory of Atomic and Solid State Physics, Cornell University, Ithaca, NY, 14853, USA}
\author{Naoki Kikugawa}
\affiliation{National Institute for Materials Science, Tsukuba, Ibaraki 305-0003, Japan}
\author{Dmitry A. Sokolov}
\affiliation{Max Planck Institute for Chemical Physics of Solids, N{\"o}thnitzer Stra{\ss}e 40, 01187 Dresden, Germany}
\author{Andrew P. Mackenzie}
\affiliation{Max Planck Institute for Chemical Physics of Solids, N{\"o}thnitzer Stra{\ss}e 40, 01187 Dresden, Germany}
\affiliation{Scottish Universities Physics Alliance, School of Physics and Astronomy, University of St.\ Andrews, St.\ Andrews KY16 9SS, United Kingdom}
\author{Clifford W. Hicks}
\email{hicks@cpfs.mpg.de}
\affiliation{Max Planck Institute for Chemical Physics of Solids, N{\"o}thnitzer Stra{\ss}e 40, 01187 Dresden, Germany}
\author{Igor I. Mazin}
\affiliation{Center for Computational Materials Science, U.S. Naval Research Laboratory, Washington, DC 20375, USA}
\date{\today}

\begin{abstract}
Uniaxial pressure applied along an Ru\nobreakdash-Ru bond direction induces an elliptical distortion of the largest Fermi surface of \SRO{}, eventually causing a Fermi surface topological transition, also known as a Lifshitz transition, into an open Fermi surface.
There are various anomalies in low-temperature properties associated with this transition, including maxima in the superconducting critical temperature and in resistivity.
In the present paper, we report new measurements, employing new uniaxial stress apparatus and new measurements of the low-temperature elastic moduli, of the strain at which this Lifshitz transition occurs: a longitudinal strain $\varepsilon_{xx}$ of $(-0.44\pm0.06)\cdot10^{-2}$, which corresponds to a B$_{1g}$ strain $\varepsilon_{xx} - \varepsilon_{yy}$ of $(-0.66\pm0.09)\cdot10^{-2}$.
This is considerably smaller than the strain corresponding to a Lifshitz transition in density functional theory calculations, even if the spin-orbit coupling is taken into account.
Using dynamical mean-field theory we show that electronic correlations reduce the critical strain.
It turns out that the orbital anisotropy of the local Coulomb interaction on the Ru site is furthermore important to bring this critical strain close to the experimental number, and thus well into the experimentally accessible range of strains.
\end{abstract} 

\maketitle

{\bf Introduction.}
Uniaxial stress has proved to be a valuable tool for investigating the unconventional superconductor \SRO{}. 
Stress applied along an Ru\nobreakdash-Ru bond, whether compressive or tensile, causes the critical temperature $T_{\text{c}}$ to increase~\cite{Hicks14}.
Under a stronger stress, $T_{\text{c}}$ passes through a pronounced peak, reaching a value of more than twice $T_{\text{c}}$ of unstressed \SRO{}~\cite{Steppke17}.
A plethora of other nontrivial renormalizations have been subsequently detected at or near the same critical strain.
The upper critical field $H_{\text{c2}}$ at the same strain is enhanced by a factor of approximately twenty~\cite{Steppke17}.
The low-temperature resistivity is observed to peak there, and the temperature exponent to be reduced from the Fermi liquid value to about 1.5~\cite{Barber18}.
Similar enhancement manifests itself in O Knight shift~\cite{Luo19}, consistent with increased partial density of O states on one O and an overall enhancement of ferromagnetic spin fluctuations on all sites.
There is a maximum in the heat capacity, consistent with an increase in the density of states~\cite{Li19}.
While the detailed mechanism of the $T_{\text{c}}$ and $H_{\text{c2}}$ enhancement is yet unclear, all observations strongly point to a strain-induced Lifshitz transition in the $\gamma$ Fermi surface sheet of \SRO{}, bringing a Van Hove Singularity (VHS) in the electronic density of states to the Fermi level.
In-plane anisotropic stress causes an elliptical distortion of this Fermi surface sheet, until it eventually transitions into an open Fermi surface.
The hypothesized Lifshitz transition has been directly observed by angle-resolved photoemission (ARPES), on a sample of \SRO{} mounted on a sample stage that uses differential thermal contraction to apply anisotropic strain~\cite{Sunko19}.

Throughout the measurements listed above, there is uncertainty in the strain $\varepsilon_\text{VHS}$ at which the Lifshitz transition occurs.
The range of values reported for $\varepsilon_{\text{VHS}}$, generally between $-5$ and $-7\times10^{-3}$, indicates the level of uncertainty.
(Negative values of $\varepsilon$ denote compression.)
Except for the ARPES measurement, all measurements were performed with piezoelectric-based uniaxial stress apparatus in which the applied strain was determined using a displacement sensor placed in parallel with the sample.
This means that the sensor measures the sum of displacements arising from sample strain, deformation of the epoxy holding the sample, and deformation of the apparatus overall.
The uncertainty in $\varepsilon_\text{VHS}$ arises chiefly from difficulties in subtracting off these additional contributions.
More accurate determination of $\varepsilon_\text{VHS}$ is important as a point of metrology, and for understanding the electronic structure.
With this aim, some of us recently developed a uniaxial stress apparatus that incorporates a force sensor placed in series with the sample~\cite{Barber19}.
The reading from the force sensor is, except for minor parasitic coupling, independent of displacement applied to the sample and epoxy, and therefore is a much more accurate and repeatable measure of the state of the sample.
In Ref.~\onlinecite{Barber19}, the resistivity $\rho$ of \SRO{} was found to peak, for temperature $T=5$~K, at a uniaxial stress of $\sigma_{xx}=-0.7$~GPa.

Here, we report three new results.
(1) Employing new measurements of the elastic moduli of \SRO{} and measurements of resistivity and susceptibility from a second sample, we determine that the Lifshitz transition occurs at a longitudinal strain $\varepsilon_{xx} = (-0.44\pm0.06)\cdot10^{-2}$; we label this strain value $\varepsilon_\text{VHS}$.
Taking into account Poisson's-ratio expansion along the transverse direction, it corresponds to a B$_{1g}$ strain $\varepsilon_{xx} - \varepsilon_{yy}$ of $\varepsilon_{\text{VHS, B$_{1g}$}} = (-0.66\pm0.09)\cdot10^{-2}$.
In this determination we assume that the peak in low-temperature resistivity marks the Lifshitz transition.
$T_\text{c}$ peaks at a slightly larger strain (by a few per cent), and although we have no experimental indication of which feature marks the transition, on general grounds it is more likely to be the peak in resistivity.
Scattering between a ``cold'', i.e.\ conventional Fermi surface and high-density-of-states hot spots has been analyzed in Refs.~\onlinecite{Berg18,Herman19}, and found to give a peak in resistivity and slower temperature dependence than $T^2$ at the Lifshitz transition.
We note that the peak in density of states at the Lifshitz transition is actually rather modest compared with that of the cold Fermi surfaces of \SRO{}~\cite{Luo19}, and that spin-fluctuation scattering, which is directly sensitive to the density of states via the Stoner renormalization, may be the larger contribution to the peak in resistivity; strong enhancement of spin fluctuations can give the observed $T^{1.5}$ temperature dependence~\cite{Moriya85}.
In both cases, however, resistivity peaks at the Lifshitz transition, while $T_{\text{c}}$ also involves a pairing interaction, which does not necessarily peak at $\varepsilon_{\text{VHS}}$.
(2) Although results for $\varepsilon_{\text{VHS}}$ from electronic structure calculations based on density functional theory (DFT) are sensitive to the precise calculation method, $\varepsilon _{\text{VHS}}$ is consistently overestimated in these calculations.
(3) Electronic correlations within dynamical mean-field theory (DMFT) substantially reduce $\varepsilon _{\text{VHS}}$.
Furthermore, the shape of the Fermi surface and thus also the critical strain subtly depend on the anisotropy of the local Coulomb interaction on the Ru site.

\begin{figure}[pt]
\centering
\includegraphics{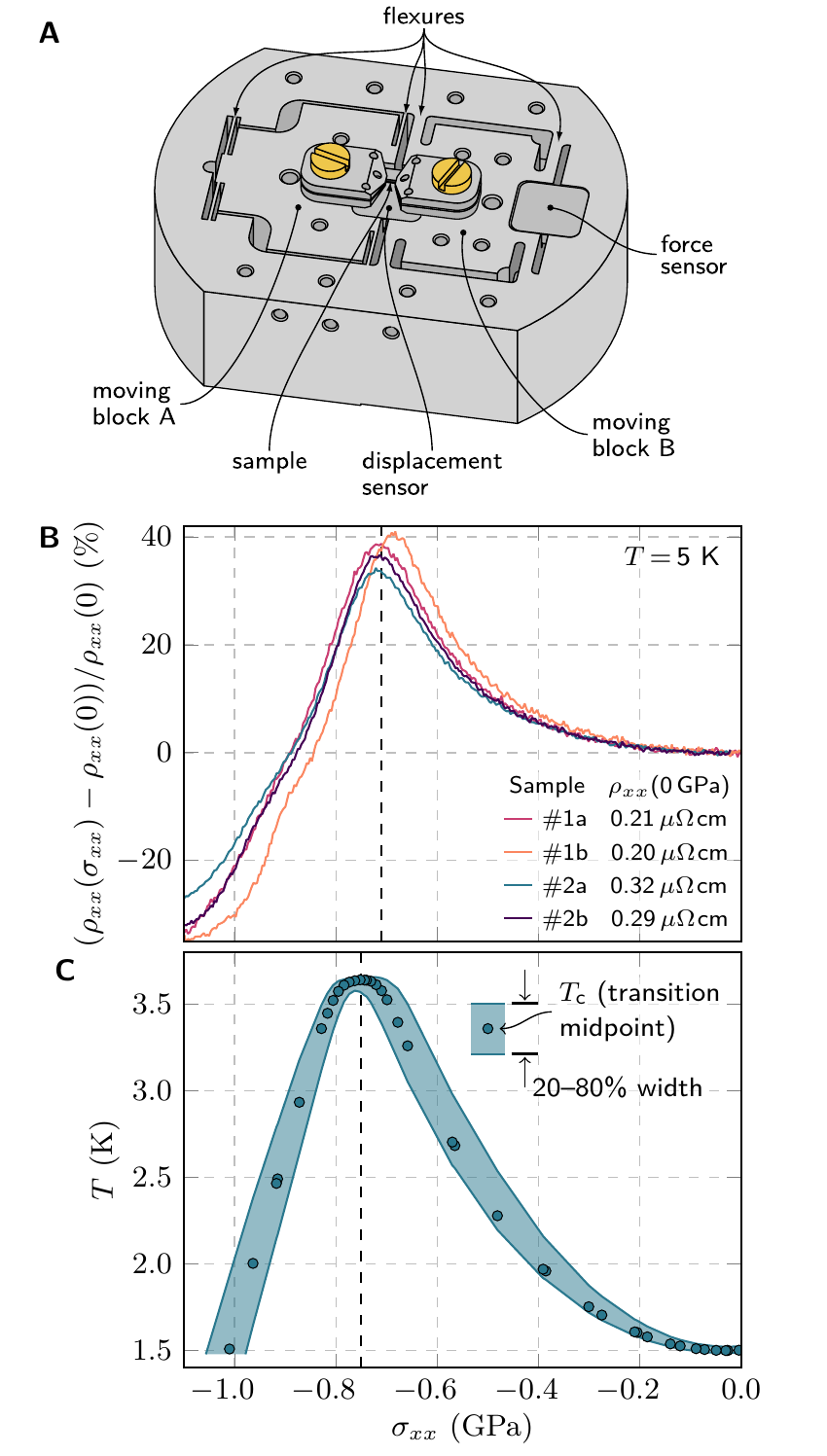}
\caption{\label{fig1} (a) Schematic of the uniaxial stress cell incorporating both a displacement and force sensor.
Both sensors are parallel-plate capacitors, placed underneath protective covers.
(b) The resistivity of two different samples as a function of applied stress.
For redundancy, two pairs of voltage contacts were attached to each sample; the letters a and b refer to pairs of voltage contacts on opposite sides of the same sample.
(c) $T_{\text{c}}$ of the second sample, determined through magnetic susceptibility measurements and measured during the same cool-down as resistivity, as a function of applied stress.}
\end{figure}

{\bf Experimental Results.}
The stress apparatus is illustrated in Fig.~\ref{fig1}A, and full details of its design are given in Ref.~\onlinecite{Barber19}.
Piezoelectric actuators underneath the device apply a displacement to moving block A, to which one end of the sample is attached.
The other end is secured to moving block B, which is secured to the outer frame through thick titanium bars, labelled flexures in the figure.
Under force applied from the sample, these flexures bend slightly, allowing block B to move.
A parallel-plate capacitive sensor between the block and outer frame measures the displacement of the block from its zero-force position, $\Delta x$.
The force on the sample is then determined as $F = k \Delta x$, where $k$ is the spring constant of the titanium flexures.
$k$ was calculated to be 20~N/$\mu$m, and measured at room temperature to be 19~$\pm$~2~N/$\mu$m.
The Young's modulus of titanium increases by $14\pm1$\% between 300 and 0~K\cite{Ekin}, so we take the low-temperature value of $k$ to be $22\pm2$~N/$\mu$m.
(All experimental error bars are 2$\sigma$, where $\sigma$ is one standard deviation.) 

The relationship between the capacitance of the force sensor $C$ and $\Delta x$ was measured both at room and cryogenic temperatures using a fibre-based interferometer.
The results are well-fitted with a parallel-plate form: $C = \epsilon_0 A / (x_0 + \Delta x) + C_\text{offset}$, where $A = 3.76$~mm$^{2}$ is the area of the capacitor plates and $x_0$ is the plate spacing at zero force.
$C_\text{offset}$ is an offset due to stray capacitance within the cell, and was measured to be $0.29\pm0.05$~pF.
$C$ was generally around 2~pF.
To determine $\Delta x$ it is also necessary to know the zero-force reading of the force sensor, $C_{0}$.
Because $C_{0}$ varies with temperature and drifts over time with repeated thermal cycling, we measured $C_{0}$ in situ, by compressing the sample until it fractured completely and then pulling apart the remnants, such that no force could have been applied through the sample to the force block.
$\Delta x$ is then calculated as $\Delta x = \epsilon_{0} A ((C-C_{\text{offset}})^{-1} - (C_{0}-C_{\text{offset}})^{-1})$.

Two samples were measured.
The resistivity data for sample~1 are also reported in Ref.~\onlinecite{Barber19}.
For sample~2, both resistivity and magnetic susceptibility were measured; susceptibility was determined through measurement of the mutual inductance between two concentric coils, the larger with diameter $800$~$\mu$m, placed on top of the sample.
Sample~1 had dimensions $L_{\text{exp}} \times W \times H$, where $L_{\text{exp}}$ is the length of the exposed central portion of the sample, $W$ the width, and $H$ the height along the $c$ axis, of $1.0\times0.37\times0.10$~mm, and sample~2, $1.0\times0.34\times0.08$~mm.
Both samples were cut from the same original rod, however were found to have slightly different residual resistivities.
The resistivity at 5~K for both samples is presented in Fig.~\ref{fig1}B.
The peak in resistivity occurs at a stress $\sigma_{xx} = -0.71\pm0.08$~GPa, where the dominant source of error is the 10\% uncertainty on $k$.

To convert the measured stress to strain, new resonant ultrasound spectroscopy measurements were performed to determine the low-temperature elastic moduli of \SRO{}.
The room-temperature Young's modulus for compression along a $\langle 100 \rangle$ direction, $Y_{100} \equiv \sigma_{xx}/\varepsilon_{xx}$ was found to be 180~GPa, as compared with 176~GPa reported in Ref.~\onlinecite{Paglione02}.
Measurements were extended to low temperature, and $Y_{100}$ was found, unusually, to decrease with cooling.
At 4~K, $Y_{100}$ is $160.2\pm0.8$~GPa, where the error bars were determined by a bootstrap method.
The in-plane Poisson's ratio for $\langle 100 \rangle$ compression is $-\varepsilon_{yy}/\varepsilon_{xx} = 0.508\pm0.006$, and the out-of-plane Poisson's ratio $-\varepsilon_{zz}/\varepsilon_{xx} = 0.163\pm0.004$.
These parameters yield $\varepsilon_\text{VHS} = (-0.44\pm0.06)\cdot10^{-2}$. 

$T_{\text{c}}$ of sample 2, shown in Fig.~\ref{fig1}C, peaks at a strain about 5\% larger than that at which resistivity peaks. In previous measurements on two further samples, the difference was found to be 11 and 12\%~\cite{Barber18}.

{\bf Calculation results.} 
We note first a technical point.
The low-temperature elastic moduli were not available at the time the calculations were performed, and therefore the room-temperature Poisson's ratios were employed.
The room-temperature in-plane Poisson's ratio is 0.39~\cite{Paglione02}, against, as stated above, 0.51 at low temperature.
The electronic structure of \SRO{} is much more sensitive to the B$_{1g}$ strain $\varepsilon_{xx} - \varepsilon_{yy}$ than the A$_{1g}$ strain $\varepsilon_{xx} + \varepsilon_{yy}$~\cite{Sunko19}.
In the theory portion of this paper we quote strains as calculated, and for comparison with experiment they should be scaled by 1.39/1.51, which matches the B$_{1g}$ strain. This is not a large enough factor to alter any of our conclusions.

We start with results of straightforward DFT calculations with and without SOC (technical details are described in the Appendix).
In Fig.~\ref{fig2} the calculated band structure near the Y point is shown.
Three points are worth noting.
First, at the Y point there are two unoccupied states: $xy$ and $yz$.
In the simplest tight binding model, they do not hybridize, and we know that in full electronic structure calculations this hybridization is indeed very weak.
However, including SOC triggers interaction between the two states.
Indeed, while the matrix element of the $L_{z}$ operator between the two states is 0, that of the $L_{\pm}$ operators is $\pm$1.
This pushes the lowest state ($xy)$ down by an amount of the order of $\lambda^{2}/(E_{yz}-E_{xy}),$ where $\lambda$ is the SOC constant, thus reducing the critical strain, an effect already noticed in Ref.~\onlinecite{Steppke17}.
Second, a technical but very important point: $\varepsilon_\text{VHS}$ is sensitive to the Fermi energy on a scale of a few meV, and a fine-scale discretization of the Brillouin zone is required to achieve convergence on this level.
We increased the number of $k$-points until convergence on this scale was achieved.
Third, the exact position of the Van Hove singularity $E_{xy} - E_{\text{F}}$ in the unstrained structure, and correspondingly the calculated $\varepsilon_\text{VHS}$, depends on the flavor of the density functional used, with the gradient-corrected functional (GGA) giving a smaller $E_{xy} - E_{\text{F}}$ than the local density approximation (LDA).

\begin{figure}[pt]
\centering
\includegraphics[width=0.85\columnwidth]{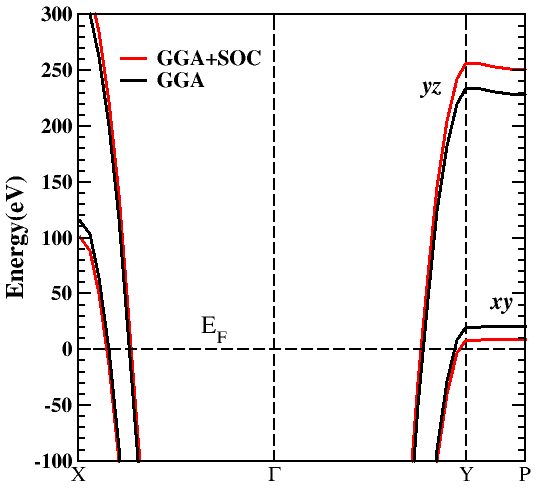}
\caption{\label{fig2} Band structure calculated within GGA~\cite{Perdew96} and GGA+SOC for $\varepsilon_{xx}=-0.75$\% using the full-potential WIEN2k code~\cite{Wien2k}.}
\end{figure}

\begin{figure}[pt]
\centering
\includegraphics[width=\columnwidth]{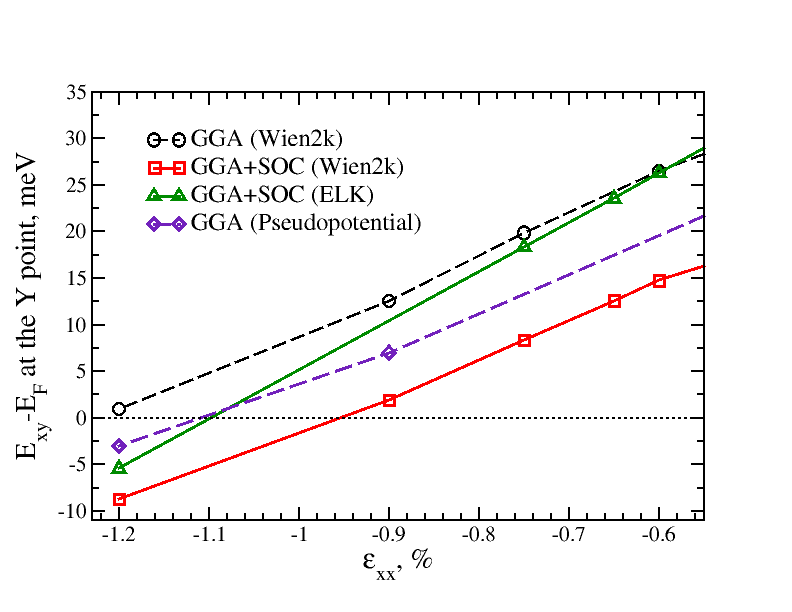}
\caption{\label{fig3} Energy of the $xy$ band at the $Y$ point as a function of $\varepsilon_{xx}$, as obtained from various codes and approximations.
Solid lines: GGA+SOC; dashed: GGA\cite{Perdew96}.
The Van Hove strain $\varepsilon_\text{VHS}$ is the strain where $E_{xy}-E_{\text{F}}$ crosses zero.}
\end{figure}

In order to further assess the sensitivity of this parameter to computational details, we have compared calculations using two all-electron augmented plane wave methods: WIEN2k~\cite{Wien2k} and ELK~\cite{ELK}.
Calculations were performed using the optimized structures reported in Ref.~\onlinecite{Steppke17} at each strain.
First, we observe that GGA (PBE~\cite{Perdew96}) indeed gives $\varepsilon_{\text{VHS}}$ smaller by $10^{-3}$ than LDA (PW~\cite{Perdew92}).
Second, adding SOC substantially reduces, as expected, the calculated Van Hove strain, which without SOC comes to at least $-1.1\times10^{-2}$: more than twice as much as the measurement, see Fig.~\ref{fig3}.
This is also higher than the value calculated in Ref.~\onlinecite{Steppke17}, which may be due to the fact that calculations employed a different basis set, namely local orbitals, and/or that it used the Dirac rather than the Pauli equation to describe relativistic effects.
In all cases, one may conclude that DFT+SOC does not fully reproduce $\varepsilon_{\text{VHS}}$.

This is not surprising per se, because it is well known that correlation effects may dramatically renormalize effective masses and crystal fields.
This is liable to alter the relative occupations of different bands and affect both the $E_{yz}-E_{xy}$ separation and the position of the Fermi level with respect to the $xy$ band, and thus the $\gamma$ sheet of the Fermi surface.
Indeed by using straightforward DFT+DMFT calculations, Zhang \textit{et al.}~\cite{Zhang16} found that for unstrained \SRO{} the Van Hove singularity shifts much closer to the Fermi level, thereby stretching the $\gamma$ sheet towards the X and Y points.
In fact, this deteriorates the agreement of the Fermi surface geometry with experiment.
It was shown, however, in the same work, that using orbital-dependent (\textit{i.e.,} anisotropic) local Coulomb parameters for the Ru-$t_{2g}$ manifold (as suggested by constrained-random-phase-approximation calculation of these parameters~\cite{Vaugier12}) radically improves the agreement with experiment.
We will show below that accounting for this anisotropy is even more important under strain.

With this in mind, we have performed DFT+DMFT calculations using the rotationally invariant slave-boson (RISB)~\cite{Li89,Lechermann07} scheme as impurity solver to reveal the effect of the interplay between band structure, spin-orbit coupling and electron correlations on the critical strain.
In line with Ref.~\onlinecite{Zhang16} as well as other works~\cite{Gorelov10,Facio18}, we choose $U=3.1$~eV and $J_{\mathrm{H}}=0.7$~eV for the isotropic Coulomb parameters in the correlated subspace spanned by the three Wannier $t_{2g}$ orbitals of Ru$(4d)$ character.
The SOC constant $\lambda$ is set to $0.09$~eV, consistent with previous DFT examinations~\cite{Haverkort08}.

As a first result, we confirm previous findings for the $k$-resolved spectral function of \SRO{} with SOC and local Coulomb interactions: electronic correlations cause substantial band renormalization.
Second, we find that correlations enhance an effect of SOC of shifting apart different Fermi-surface sheets, leading also to a slight redistribution of electrons across those sheets.
The increased separation between the Fermi surface sheets may give an impression that correlations have strengthened SOC, however far from $E_{\text{F}}$ the SOC-driven avoided crossings are not strongly affected by correlations~\cite{Behrmann12,Kim18,Facio18}.
In other words, the underlying SOC is not substantially altered.
In our calculations, at the $\Gamma$ point the splitting between the $\beta$ and $\gamma$ bands, which at $\Gamma$ are both well below $E_F$, decreases from 91 to 78~meV when interactions are introduced.

Crucially for the present discussion, correlations beyond DFT lead to growth of the $\gamma$ sheet relative to the $\alpha$ and $\beta$ sheets.
As a result, $\varepsilon_\text{VHS}$ is considerably reduced compared to the DFT value, in line with the previous work by Facio \textsl{et al.}~\cite{Facio18}.
However for the chosen set of local Coulomb parameters and SOC, the critical strain is shifted essentially to zero.
In essence, $\varepsilon_{\text{VHS}}$ obtained from DFT is \emph{too large}, while from conventional DFT+DMFT it is \emph{too small}.

\begin{figure}[pb]
\centering
\includegraphics[width=\columnwidth]{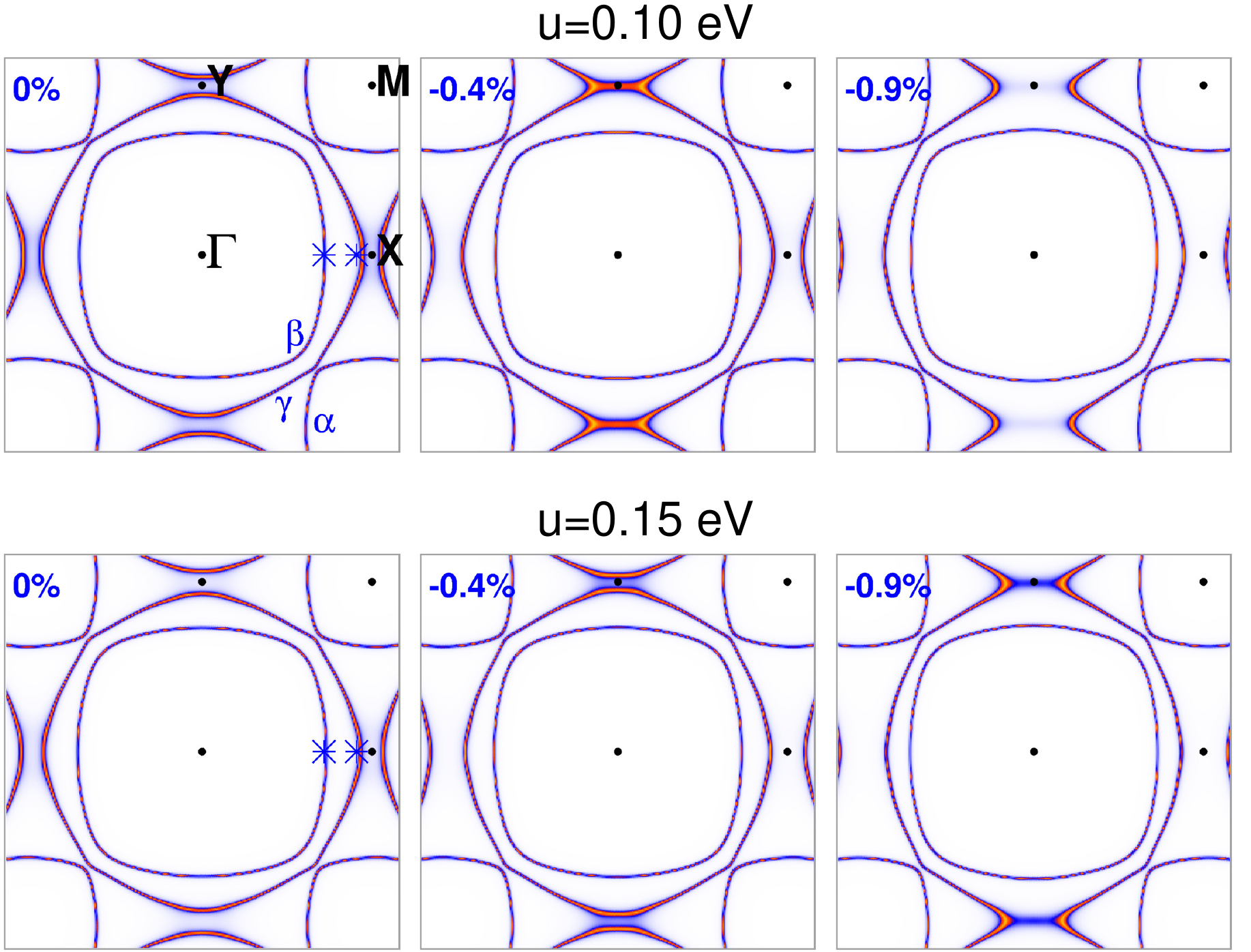}
\caption{\label{fig4} Evolution of the correlated Fermi surface with $\alpha$, $\beta$ and $\gamma$ sheets for the local Coulomb anisotropies $u=0.10$~eV (top) and $u=0.15$~eV (bottom) from zero strain to $\varepsilon_{xx} = -0.4$\% and $-0.9$\%.
Blue asterisks mark the experimental Fermi crossings along $\Gamma$-X for the unstrained compound~\cite{Tamai18}.
The strain-induced Lifshitz transition takes place close to the Y point in the Brillouin zone.}
\end{figure}

Stimulated by the findings of the importance of anisotropy in the local Coulomb parameters~\cite{Zhang16}, we introduced orbital differentiation within the Hubbard-U matrix.
The $xy$ Wannier function is larger in spatial extent compared to $\{xz,yz\}$, making it more sensitive to the Coulomb repulsion among the O$(2p)$ electrons~\cite{Mazin97-1,Mazin97-2}.
Therefore, a larger $U_{xy,xy}$ matrix element is expected, consistent with Ref.~\onlinecite{Vaugier12}.
We treat the intra-orbital Coulumb anisotropy $u=U_{xy,xy}-U_{\{xz,yz\},\{xz,yz\}}$ as a parameter.
The second anisotropic term, the inter-orbital Coulomb anisotropy $u^{\prime}=U_{xy,\{xz,yz\}}-U_{\{xz,yz\},\{yz,xz\}}$ is set to $u^{\prime}=u/3$, following Ref.~\onlinecite{Zhang16}.
Hence, the different orbital-dependent local Hubbard interactions read 
\begin{equation}
\begin{split}
U_{xy,xy}&=U+\frac{u}{2}\ ,\\
U_{\{xz,yz\},\{xz,yz\}}&=U-\frac{u}{2}\ ,\\
U_{xy,\{xz,yz\}}&=U-2J_{\text{H}}+\frac{u'}{2}\ ,\\
U_{\{xz,yz\},\{yz,xz\}}&=U-2J_{\text{H}}-\frac{u'}{2}\ .
\end{split}
\end{equation}
Note of course that spin-flip and pair-hopping account still for a further isotropic $J_{\text{H}}$ contribution in the local Hamiltonian.
In essence, the stronger Coulomb interaction for electrons in the $\gamma$ sheet relative to $\alpha$ and $\beta$ raises the energy of states in this sheet, which partially unwinds the above-mentioned correlation-driven transfer of particles from the $\alpha$ and $\beta$ sheets to the $\gamma$ sheet.

Figure~\ref{fig4} depicts the strain evolution of the Fermi surface for $u=0.10$~eV and $u=0.15$~eV.
Both values give a very good agreement with the experimental Fermi surface at zero strain~\cite{Tamai18}, as indicated by the blue asterisks in the figure.
It is seen that the strain-induced Lifshitz transition at Y is shifted to larger strain values for larger $u$.
A linear interpolation of the calculated $E_{xy}-E_{\text{F}}$ at strains $\varepsilon_{xx} = \{$0, $-0.4\cdot10^{-2}$, $-0.9\cdot10^{-2}\}$ yields $\varepsilon_\text{VHS} = -0.38\cdot10^{-2}$ for
$u=0.10$~eV and $-0.70\cdot10^{-2}$ for $u=0.15$~eV.
This demonstrates the sensitivity to local Coulomb anisotropy, and also shows its potential of driving the theoretical critical strain closer to the experimental value. 

In agreement with the strain ARPES data of Ref.~\onlinecite{Sunko19}, the DFT+DMFT calculations here indicate no substantial strain-induced redistribution of carriers between the Fermi surface sheets.
For both $u = 0.10$ and 0.15~meV, the $\beta$ sheet has between 2 and 3\% more area at $\varepsilon_{xx} = -0.9\cdot10^{-2}$ than at zero strain.
This result should be verified with other impurity solvers, however it is anyway well below the resolution of the ARPES data in Ref.~\onlinecite{Sunko19}.

We note that a lower isotropic $U=2.3$~eV, and $J_{\text{H}}=0.4$~ eV~\cite{Kim18,Tamai18} together with same $u,u^{\prime}$ as above also yields a reasonable match to the experimental Fermi surface at zero strain.
However this choice of parameters gives both substantially larger $\varepsilon_\text{VHS}$ than what is observed, and smaller averaged band renormalization than the experimental range of $m^{\ast}/m\sim 2.5$--4.4~\cite{Puchkov98,Tamai18}.
Our choice reproduces the lower limit of this range.
Note that in the RISB solution to DMFT one typically needs to use slightly larger $U$ values to cope with the correlation strength than those utilized, for instance, in the quantum Monte-Carlo impurity solvers.
However in any case, due to the interplay of Hund's-metal physics and Van-Hove-induced fluctuations acting on different bands, the mass renormalization displays significant band dependence~\cite{Tamai18}.

{\bf Conclusions.}
Firstly, through new measurements we have refined the experimental determination of $\varepsilon_\text{VHS}$, the longitudinal strain at which the $\gamma$ sheet is driven through a Lifshitz transition when \SRO{} is uniaxially pressurized along an Ru\nobreakdash-Ru bond direction, to $(-0.44\pm0.06)\cdot10^{-2}$.
This corresponds to $\varepsilon_{xx} - \varepsilon_{yy} = (-0.66\pm0.09)\cdot10^{-2}$.
Secondly, although there is some uncertainty in the first-principles calculation, $\varepsilon_\text{VHS}$ is found to be considerably overestimated in DFT calculations, even when SOC is included.
We find that Coulomb repulsion reduces the calculated Van Hove strain, however an isotropic Coulomb interaction strong enough to reproduce the observed band renormalizations gives $\varepsilon_{\text{VHS}}$ much smaller than observed.
Good agreement with experimental data can be achieved only by accounting for orbital anisotropy in the Coulomb interactions on the Ru site.
For $U = 3.1$~eV and $J_{\text{H}}=0.7$~eV, the experimental $\varepsilon_\text{VHS}$ is reproduced for $u \equiv U_{xy,xy}-U_{\{xz,yz\},\{xz,yz\}}\sim$0.10--0.15~eV.
This finding highlights the value of high-precision strain measurements in providing an additional constraint on theory.
They also highlight a practical benefit of correlations, in reducing the strain required to reach the Lifshitz transition, and the important observations as it is crossed.

\begin{acknowledgments}
We acknowledge the support of the Max Planck Society.
F. L. was supported by the DFG under the project LE-2446/4-1.
S. V. S. and S. L. S.'s work was supported by the Russian Ministry of Science and High Education through program AAAA-A18-118020190095-4 (``Quantum'') and contract 02.A03.21.0006, as well as UD RAS via project 18-10-2-37.
B. J. R. and S. G. were supported by the National Science Foundation under Grant No.\ DMR-1752784.
N. K. acknowledges the support from JSPS KAKENHI (No.\ JP18K04715) and JST-Mirai Program (No.\ JPMJMI18A3) in Japan.
I. I. M. was supported by ONR through the NRL basic research program.
The authors are thankful to Helge Roesner for many useful discussions, and for providing the crystal structures fully optimized as a function of strain in Ref.~\onlinecite{Steppke17}.
\end{acknowledgments}

\section*{Appendix}

Ab initio density functional calculations were performed using three different codes, two of which are based on the all-electron linearized augmented plane wave method (WIEN2k~\cite{Wien2k} and ELK~\cite{ELK}) and the third one is a mixed-basis pseudopotential scheme~\cite{mbpp_code,Lechermann02}.
The first two are among the most accurate methods (codes) present on the DFT market.
Selected points were also checked against the projector-augmented-wave method~\cite{Bloechl94} implemented in the Vienna Ab initio Simulation Package (VASP)~\cite{Kresse96}.

The convergence with respect to the size of the basis set as well as to the $k$-point mesh has been checked (in WIEN2k a mesh up to 50$\times$50$\times$50 $k$-points and $RK_{\text{max}}=8.5$ was used, and for ELK 30$\times$30$\times$30 and $RK_{\text{max}}=9.0$).
Results of the WIEN2k and ELK calculations (together with those of Ref.~\onlinecite{Steppke17}) set up an error bar for the critical strain as obtained in DFT.
It was also used as a hallmark for the pseudopotential code, which was utilized in the DFT+DMFT calculations.
The generalized-gradient approximation in the Perdew-Burke-Ernzerhof version~\cite{Perdew96} was used in these calculations for the exchange-correlation potential.
The DMFT correlated subspace is defined by the maximally-localized Wannier functions~\cite{Marzari12} for the Ru$(4d)$-$t_{2g}$ orbitals.
A three-orbital Hubbard Hamiltonian of Slater-Kanamori form, parametrized by the Hubbard $U$ and the Hund's exchange $J_{\text{H}}$, is applied in that subspace. 

The spin-orbit coupling described by $H_{\text{soc}}=\lambda /2\sum_{mm^{\prime}}\sum_{\sigma\sigma^{\prime}} c^{\dagger}_{m\sigma }\,\bm{L\cdot\sigma}\,c_{m^{\prime}\sigma^{\prime}}$ was taken into account for correlated orbitals space.
Here $m,m^{\prime}$ are orbital indices, $\sigma,\sigma^{\prime}$ mark spin projections, $\bm{L}$ are $t_{2g}$ angular-momentum matrices, and $\bm{\sigma}$ denotes the Pauli matrices.

The impurity problem was solved by the rotationally invariant slave-boson (RISB)~\cite{Li89,Lechermann07} framework at the saddle point.
The RISB electronic self-energy consists of a term linear in frequency as well as a static part, and thus has a simpler and more restricted form as a possible general local $\Sigma(\omega)$.
However, the RISB scheme (here at formal $T=0$) is still well-suited for many correlated materials problems such as ruthenates~\cite{Behrmann12,Facio18}.

For all calculations, crystal structures as optimized in Ref.~\onlinecite{Steppke17} as a function of strain were used as the structural input.

\end{document}